\documentclass[sigconf]{acmart}
\setcopyright{none} 
\renewcommand\footnotetextcopyrightpermission[1]{} 
\usepackage[utf8]{inputenc}
\usepackage[colorinlistoftodos,prependcaption,textsize=tiny]{todonotes}

\usepackage{hyperref}
\definecolor{linkcolor}{rgb}{0.65,0,0}
\definecolor{citecolor}{rgb}{0,0.4,0}
\definecolor{urlcolor}{rgb}{0,0,0.65}
\hypersetup{pdfpagemode=UseNone,pdfstartview=FitH,colorlinks=true, linkcolor=linkcolor, urlcolor=urlcolor, citecolor=citecolor}

\usepackage{makecell}
\usepackage{listings}
\usepackage{color}
\usepackage{multicol}
\usepackage{algorithmicx}
\usepackage[ruled]{algorithm}
\usepackage{algpseudocode} 
\usepackage{amsmath}
\usepackage{subfig}

\newcommand*\BitAnd{\mathrel{\&}}
\newcommand*\BitOr{\mathrel{|}}

\newcommand*\BitNeg{\ensuremath{\mathord{\sim}}}

\definecolor{dkgreen}{rgb}{0,0.6,0}
\definecolor{gray}{rgb}{0.5,0.5,0.5}
\definecolor{mauve}{rgb}{0.58,0,0.82}
\newcommand{\isdef}{\coloneqq}

\newcommand{\tool}{\emph{MicroWalk}}

\begin{document}
	\title{\tool: A Framework for Finding Side Channels in Binaries}
	

	\author{Jan Wichelmann}
	\orcid{0000-0002-5748-5462}
	\email{j.wichelmann@uni-luebeck.de}
	\affiliation[obeypunctuation=true]{%
		\institution{Universität zu Lübeck}\\
		\state{Lübeck, Germany}
	}
	\author{Ahmad Moghimi}
	\email{amoghimi@wpi.edu}
	\affiliation[obeypunctuation=true]{%
		\institution{Worcester Polytechnic Institute}\\
		\state{Worcester, MA, USA}
	}
	\author{Thomas Eisenbarth}
	\email{thomas.eisenbarth@uni-luebeck.de}
	\affiliation[obeypunctuation=true]{%
		\institution{Universität zu Lübeck \& WPI}\\
		\state{Lübeck, Germany}
	}
	\author{Berk Sunar}
	\email{sunar@wpi.edu}
	\affiliation[obeypunctuation=true]{%
		\institution{Worcester Polytechnic Institute}\\
		\state{Worcester, MA, USA}
	}
	
	\copyrightyear{2018}
	\acmYear{2018}
	\setcopyright{acmlicensed}
	\acmConference[ACSAC '18]{2018 Annual Computer Security Applications Conference}{December 3--7, 2018}{San Juan, PR, USA}
	
	\acmPrice{15.00}
	\acmDOI{10.1145/3274694.3274741}
	\acmISBN{978-1-4503-6569-7/18/12}
	
	\begin{abstract}
		Microarchitectural side channels expose unprotected software to information leakage attacks where a software adversary is able to track runtime behavior of a benign process and steal secrets such as cryptographic keys. As suggested by incremental software patches for the RSA algorithm against variants of side-channel attacks within different versions of cryptographic libraries, protecting security-critical algorithms against side channels is an intricate task. Software protections avoid leakages by operating in constant time with a uniform resource usage pattern independent of the processed secret. In this respect, automated testing and verification of software binaries for leakage-free behavior is of importance, particularly when the source code is not available. In this work, we propose a novel technique based on Dynamic Binary Instrumentation and Mutual Information Analysis to efficiently locate and quantify memory based and control-flow based microarchitectural leakages. We develop a software framework named \tool~for side-channel analysis of binaries which can be extended to support new classes of leakage. For the first time, by utilizing \tool, we perform rigorous leakage analysis of two widely-used closed-source cryptographic libraries: \emph{Intel IPP} and \emph{Microsoft CNG}. We analyze $15$ different cryptographic implementations consisting of $112$ million instructions in about $105$ minutes of CPU time. By locating previously unknown leakages in hardened implementations, our results suggest that \tool~can efficiently find microarchitectural leakages in software binaries. 
		
	\end{abstract}

	\maketitle

	\section{Introduction}
	Side-channel attacks exploit information leakage through physical behavior of computing devices. The physical behavior depends on the processed data. The resulting data-dependent patterns in physical signals such as power consumption, electromagnetic emanations or timing behavior can be analyzed to extract secrets such as cryptographic keys~\cite{kocher1996timing,brumley2005remote,mangard2008power,genkin2015stealing}. Despite the physical proximity requirement for most physical attacks, there exist remotely exploitable side channels such as 
	microarchitectural attacks~\cite{ge2016survey}. 
	
	Microarchitectural attacks exploit shared hardware features such as cache~\cite{bernstein2005cache,percival2005cache,osvik2006cache}, branch prediction unit (BPU)~\cite{aciiccmez2007power}, memory order buffer (MOB)~\cite{moghimi2017memjam} and speculative execution engine~\cite{Kocher2018spectre} to extract secrets from a process executed \emph{on the same system}. These attacks can be mounted remotely or locally on systems where untrusted entities can execute code on a shared hardware, either because the system is shared or untrusted code is executed. Scenarios include but are not limited to cross-VM attacks in the cloud environment~\cite{irazoqui2015s,liu2015last}, drive-by JavaScript trojans inside the browser sandbox~\cite{lipp2017practical}, attacks originating from untrusted mobile applications~\cite{lipp2016armageddon} and system-adversarial attacks against Intel Software Guard eXtensions (SGX)~\cite{moghimi2017cachezoom,van2017telling}. Microarchitectural leakage can be used to break software implementations of cryptographic schemes where the adversaries recover the secret key by combining the leaked partial information from key-dependent activities ~\cite{pereida2016make,yarom2014recovering,benger2014ooh}. These side channels can be further exploited to violate user's privacy through activity profiling~\cite{gulmezoglu2017perfweb}, or to steal user's keystrokes~\cite{gruss2015cache}. Memory protections such as Address Space Layout Randomization (ASLR) can be bypassed by exploiting microarchitectural side-channel leakages~\cite{evtyushkin2016jump}. 
	
	Defense against microarchitectural side channels have been proposed based on new hardware design~\cite{costan2016sanctum,kayaalp2017ric}, systematic mitigation~\cite{liu2016catalyst} and activity monitoring~\cite{zhang2016cloudradar,briongos2018cacheshield}. However, the most widely-used protection against microarchitectural leakage is software hardening using constant-time programming techniques~\cite{hamburg2009accelerating,brickell2006mitigating}. In this context, constant-time programming implies using microarchitectural resources in a secret-independent fashion. Therefore, timing, or trace-based leakages~\cite{disselkoen2017prime} in the hardware would not reveal any information about the secret. These techniques depend on the underlying microarchitecture and side-channel knowledge, i.e. software implementations are hardened to follow a constant-time behavior based on published attacks on the target microarchitecture. Consequently, a novel microarchitectural attack demands new changes to these software protections. While true constant-time code avoids such problems, manual verification of the software implementation for constant-time behavior is an error-prone task, and it requires extensive, and ever growing knowledge of side channels. Besides, what we observe in the source code is not always what is executed on the processor~\cite{simon2018you}, and there are leakages in the program binary that remain unobserved in the source code~\cite{kaufmann2016constant}. The state of art tools and techniques for automated finding of side-channel leakages in software binaries fall short in practice, particularly when the source code is not available. As a result, commercial cryptographic products such as \emph{Microsoft Cryptography API Next Generation (CNG)}, which is used everyday by millions of users, have never been externally audited for side-channel security.

	\subsection{Our Contribution}
	We propose a leakage detection technique, and develop a framework named \tool~to locate leakages within software binaries. We apply \tool~to analyze two commercial closed-source cryptographic libraries hardened toward constant-time protections and report previously unknown vulnerabilities, in summary:
	\begin{itemize}
		\item We propose a technique based on Dynamic Binary Instrumentation (DBI) and Mutual Information (MI) Analysis to locate memory based and control-flow based microarchitectural leakages in software binaries. 
		\item We develop the \tool~framework to perform automated leakage testing and quantification based on our technique. Our framework can be extended to locate other and new types of microarchitectural leakages.
		\item We demonstrate the ease-of-use of \tool~by showing how it significantly eases the analysis of binary code even in cases where source code is not accessible to the analyst.
		\item We apply \tool~to cryptographic schemes implemented in \emph{Microsoft CNG} and \emph{Intel IPP}, which are both widely used, yet closed source crypto libraries. Our results include previously unknown leakages in these libraries.
		\item We perform analysis and quantification of the critical leakages, and discuss the security impact of these leakages on the relevant cryptographic schemes.
	\end{itemize}
	
	\subsection{Analysis Setup and Targeted Software}
	Our machine for analysis is a Dell XPS 8920 machine with Intel(R) Core i7-7700 processor, 16\,GB of RAM and a traditional hard disk drive running \emph{Microsoft Windows 10}. The \tool~Framework uses \emph{Pin} v3.6 as the DBI backend, and \emph{IDA Pro} v6.95 for binary visualization and leakage analysis. The tested cryptographic modules are \emph{Microsoft bcryptprimitives.dll} v10.0.17134.1 as part of \emph{Microsoft CNG}, and \emph{Intel IPP} v2018.2.185.

	\section{Background}
	
	\subsection{Dynamic Binary Instrumentation}
	Dynamic program analysis is more accurate compared to static analysis due to availability of real system states and data~\cite{nethercote2004dynamic}. Dynamic analysis requires instrumentation of the program binary, and it analyzes the program when it executes. The instrumentation code is added to the program binary without changing the normal logic and execution flow of the program under analysis, and it contains minimal instructions and subroutines for collecting metadata and measurements. The instrumentation code and the instrumented code execute at the same time following each other. Indeed, adding instrumentation is easier during the compilation phase and when the source code is available~\cite{lattner2004llvm}, but source code is not always available, and the analysis would not be as accurate due to compiler transformations. Thanks to Dynamic Binary Instrumentation (DBI) frameworks such as Pin~\cite{luk2005pin}, it is possible to instrument program binaries without source code. 
	
\begin{figure}[tp]
		\includegraphics[width=.9\columnwidth]{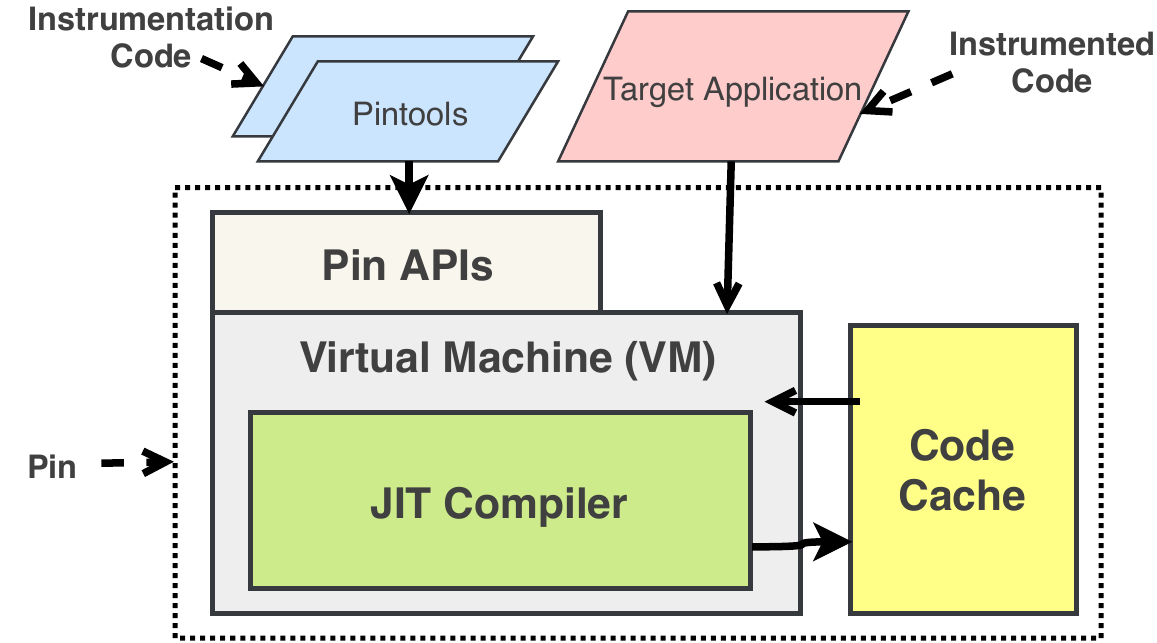}
		\caption{Pin: The JIT compiler combines application and instrumentation codes, and it stores the transformed binary in code cache. The virtual machine maintains and tracks program states, while it executes from the code cache.}
		\label{fig:pin}
		
	\end{figure}

	\emph{Pin} is a DBI framework based on just-in-time (JIT) compilation. In general, JIT compilers transform a source language to executable binary instructions at runtime. ~\autoref{fig:pin} shows how an embedded JIT engine is part of \emph{Pin} to recompile the binary instructions at runtime and combine the program's instruction with instrumentation codes, named \emph{Pintools}. To avoid the performance pitfall of JIT compilation, \emph{Pin} uses a \texttt{code cache} that stores the combined code, and re-execution of the same basic blocks occur from the \texttt{code cache}. Binary instrumentation using \emph{Pintools} gives us an easy to use interface to collect runtime metadata about program states such as the accessed memory addresses, targets of indirect branches and memory allocations. \emph{Pin} makes sure the instrumentation is transparent, i.e., it preserves the original application behavior~\cite{luk2005pin}. These events can be measures as accurate as they occur on the OS and the processor and as it would be an uninstrumented execution. In terms of microarchitectural analysis, we can observe the program behavior and resource usage as they appear on the hardware, and this gives us the ability to model a known microarchitectural leakage based on the observation of states from a real system.
	
	\subsection{Microarchitectural Leakage} \label{sec:micro}
	Modern microarchitectures feature various shared resources, and these resources are distributed among malicious and benign processes with different permissions. A malicious process, sharing the same hardware, can cause resource contention with a victim and measure the timing of either the victim or herself to learn about the victim's runtime. In a cache attack, the adversary accesses the same cache set that the victim's security-critical memory accesses are mapped to, and she measures the memory accesses' timing. A slow memory access reveals some information about the address bits of the victim's memory access. As motivated by cache attacks on AES~\cite{osvik2006cache,bernstein2005cache}, knowledge of secret-dependent memory accesses such as S-Box operations leaks information about the internal runtime state, and this information can be used for cryptanalysis and secret key recovery. In cache attacks, the size of each cache block is 64\,B which stops adversaries from gaining information about the $\log_2(64)=6$ least significant address bits. While some constant-time software countermeasures assume that the adversary cannot leak these bits, there are microarchitectural attacks on \texttt{cache banks} and \texttt{MOB} that leak beyond this assumption~\cite{yarom2017cachebleed,moghimi2017memjam}. In this work, we consider all secret-dependent memory accesses and treat them as memory-based leakages disregarding their spatial resolution. 
	
\begin{figure}[tp]
		\centering
		\begin{minipage}{.4\columnwidth}
			\begin{lstlisting}[label={lst:xxxx}]
			// Square and Multiply 
			//     Pseudocode
			while(i>0){
			r = (r * r) % n;
			if( key[--i] == 1 ){
			r = (r * x) % n;
			}	
			}            
			\end{lstlisting}
		\end{minipage}\hfil
		\begin{minipage}{.5\columnwidth}
			\includegraphics[width=\columnwidth]{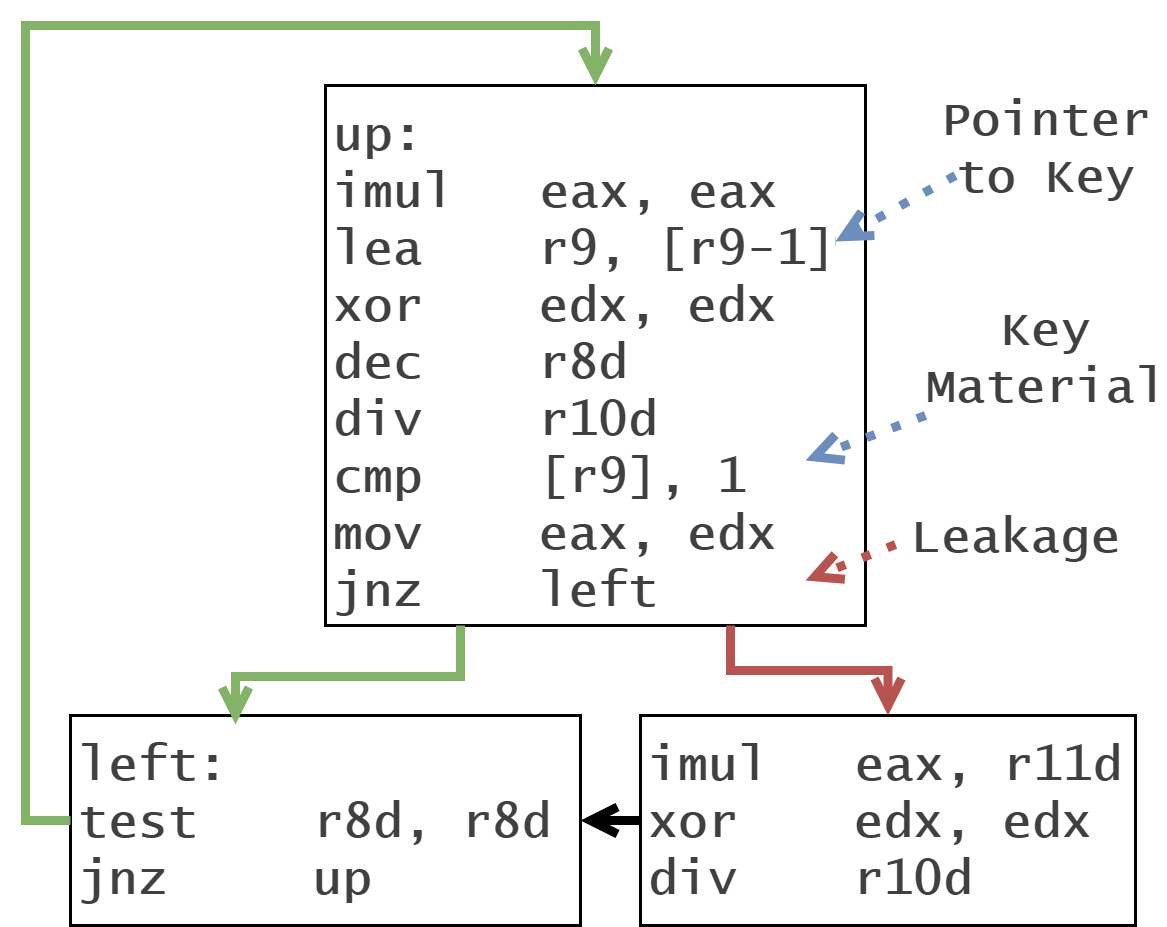}
		\end{minipage}
		\caption{Montgomery \texttt{Square and Multiply} operations can leak information about the secret exponent. While \texttt{r9} points to the exponent in memory, comparison of a value from the exponent determines if the left jump should occur which leaves a key-dependent microarchitectural footprint.}
		\label{fig:secretbranch}
	\end{figure}

	Memory operations are not the only source of leakage. A conditional statement, or a processing loop that depends on a secret to choose an execution path can leak information about the secret. Each unique execution path operates on a different set of instructions, and it consumes the shared resources uniquely. Shared resources such as \texttt{instruction cache} and \texttt{BPU} leak information about the state of branches~\cite{aciiccmez2010new,aciiccmez2007predicting}. \autoref{fig:secretbranch} resembles a classical side-channel leakage in RSA Montgomery modular exponentiation. This algorithm processes a secret exponent one bit at a time, and it performs an additional arithmetic operation when the secret bit is one. An adversary who is able to track the execution of the left branch is able to determine the secret value that affected the conditional jump decision. We treat all the attacks that are triggered due to secret-dependent branches as control-flow based attacks.
	
	\subsection{Mutual Information Analysis}
	Mutual information (MI) measures the mutual dependence of two random variables, and it can be used to quantify the average amount of obtainable information about one variable through observation of the second variable~\cite{guiau1977information}. Mutual information using Shannon entropy is defined as
	\begin{equation*}
	I(X,Y)=\sum_{x\in X}\sum_{y\in Y}p(x,y)\log_2\left(\frac{p(x,y)}{p(x)p(y)}\right),
	\end{equation*}
	
	where $p(x)$ and $p(y)$ are the probability distributions of random variables $X$ and $Y$ respectively. $p(x,y)$ is the joint probability of $X$ and $Y$, and $I(X,Y)$ tells us the average amount of dependent information in bits\footnote{$\log_2$ measures the MI in \emph{bit} unit.} between the variables $X$ and $Y$. MI has been utilized to quantify side-channel security~\cite{standaert2009unified,zhang2014new,bayrak2015automatic,Irazoqui16Leak}, or to mount side-channel attacks~\cite{gierlichs2008mutual}. Redefining MI in the side-channel context, we can define variable $X$ as the secret and variable $Y$ as an internal physical state of a system leaked through a side channel. $I(X,Y)$ will measure the average amount of leakage from secret $X$, through observing the side-channel information $Y$. 
	
	\subsection{Signing Algorithms} \label{sec:signing}
	\subsubsection{DSA}
	\emph{Digital Signing Algorithm (DSA)}~\cite{pub1993digital} is a signature scheme based on the discrete logarithm problem (DLP)~\cite{kevin1990discrete}. Choosing a prime $p$, another prime $q$ divisor of $p-1$, the group generator $g$, a secret key $x$, the public key $y=g^x \bmod p$, and the hash of the message to be signed $z$, the \emph{DSA} signing operation is defined as
	\begin{gather*}
	k\leftarrow~RANDOM\ |\ 1<k<q\\
	r = (g^k \bmod p) \bmod q,\ 
	s = k^{-1}(z + r\cdot x) \bmod q
	\end{gather*}
	where $(r, s)$ are the output signature pairs. 
	
	\subsubsection{ECDSA}
	\emph{Elliptic-Curve DSA (ECDSA)}, as an analogue of \emph{DSA}, is a signature scheme based on elliptic curves~\cite{johnson2001elliptic}, in which the subgroup of a prime $p$ is replaced by the group of points on an elliptic curve over a finite field. Choosing an elliptic curve, a point on the curve $G$, the integer order $n$ of $G$, a secret key $d_A$, the public key $Q_A=d_A \times G$, and the hash of message to be signed $z$, the ECDSA signing operation is defined as
	\begin{gather*}
	k\leftarrow~RANDOM\ |\ 1<k<n-1 \\ 
	(x_1, y_1) = k \times G\\ 
	r=x_1 \bmod n,\ 
	s=k^{-1}(z+r\cdot d_A) \bmod n
	\end{gather*}
	Both \emph{DSA} and \emph{ECDSA} use an ephemeral secret $k$ that needs to be chosen randomly for each operation. 
	
	\subsubsection{Modified Elliptic Curve Signature}
	\emph{Elliptic-Curve Nyberg-Rueppel (ECNR)}~\cite{NybergRueppel} and \emph{SM2}~\cite{sm2BaiZhang}, a standard signature scheme, are modified schemes based on \emph{ECDSA} that allow signatures with message recovery. \emph{ECNR} and \emph{SM2} are widely used, and they are both supported by \emph{Intel IPP}. Public parameters, the private/public key pair and the ephemeral secrets are chosen similar to \emph{ECDSA}. The pair $(x_1, y_1)$ is also calculated similarly, but the signature generation for \emph{ECNR} is defined as
	\begin{gather*}
	r=x_1+z,\
	s=k-r\cdot d_A,
	\end{gather*}
	and the signature generation for \emph{SM2} is defined as
	\begin{gather*}
	r=x_1+z,\ 
	s=(1+d_A)^{-1}(k-r\cdot d_A)
	\end{gather*}

	\section{\tool~Analysis Technique}\label{sec:mi_technique}
	\tool~aims to find microarchitectural leakages in software binaries. A binary implementation is vulnerable to microarchitectural side-channel attacks when there is a dependency between a secret and internal computation states observable through the side channels. We expose such relationships and quantify the amount of observable leakage in these implementations. This helps security analysts \textbf{1)} to reveal whether an implementation has leakages, \textbf{2)} to locate the exact location of each leakage in the binary, and \textbf{3)} to measure the dependency between the secret and the internal state, i.e., it can give some confidence value on the severity of the leakage. In contrast to side-channel analysis model, we are able to perform this analysis in a white-box model. 
	
	\subsection{Leakage Analysis Model}
\begin{figure}[tp]
		\includegraphics[width=.87\columnwidth]{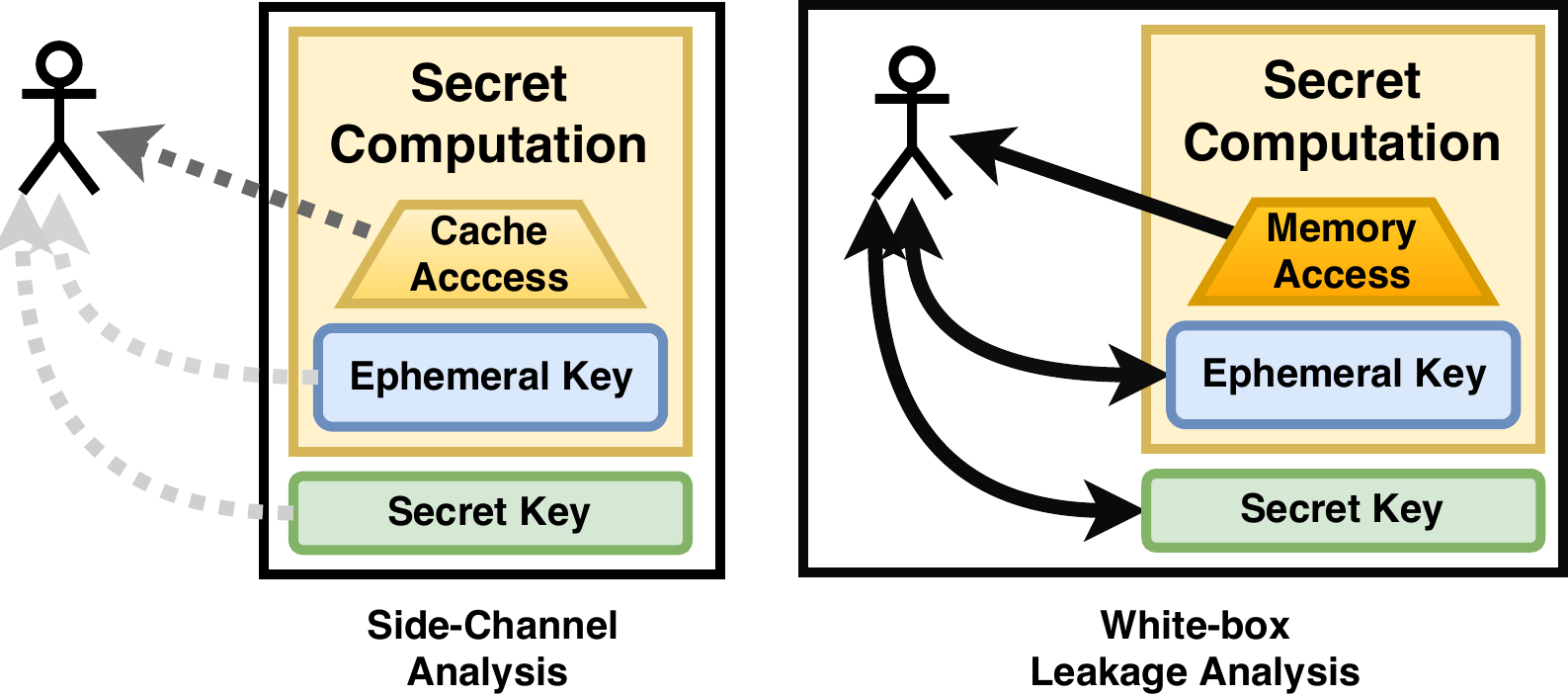}
		\caption{Left: side-channel analyst finds relationship between a real leakage such as cache access pattern and secrets such as cryptographic keys. Right: \tool~follows a white-box model where the security analyst has full access to runtime states such as memory accesses, and she can find dependencies between arbitrary secrets and internal states.}
		\label{fig:analysismodel}
		\vspace{-1.5ex}
	\end{figure}

	We assume a strong adversary with full access to runtime events such as memory accesses, execution path and even register values. Further, the adversary can choose and modify any secret input of the system. This strong adversary can define any 
	internal computation state such as addresses of memory accesses and register values as a potential leakage vector, based on her knowledge of a category of side-channel attacks, e.g., memory based attacks (\autoref{sec:micro}). The adversary executes the system under her full control, and feeds the system with arbitrary secrets while collecting runtime traces for the defined leakage vector. ~\autoref{fig:analysismodel} compares our leakage analysis model with the side-channel analysis model. As an example, if we try to analyze a binary implementation of AES, we need to define certain operations as our leakage vectors. Based on cache attacks, an adversary defines memory accesses as a leakage vector, and she collects all memory accesses during the execution of AES using arbitrary secret keys. If there is a dependency between different secret keys and the variation of memory accesses, the adversary can locate which instructions relate to any secret-dependent memory accesses, and identify potential leakages. 
	
	\subsection{Capturing Internal States}
	
	We choose two common sources of leakage as our leakage vectors: \textbf{1)} execution path and \textbf{2)} memory accesses. A true constant-time implementation follows a linear execution path for any given secret input; each time a software performs a secret-dependent conditional branch, it leaks some amount of information about the secret. Defining execution path as a leakage vector helps us to check whether for any secret input the same operations are performed. The second common source of leakage are memory accesses. A constant-time implementation should follow a secret-independent memory access pattern. If, for example, an implementation of a cryptographic algorithm does key-dependent table look ups which can be exploited by measuring cache timings, an attacker will be able to extract parts of the secret key; we ensure that memory accesses are either invariant, or at least uncorrelated to the input (e.g. blinding in RSA~\cite{kocher1996timing}). 
	
	To be able to detect these two types of leakages, we need to collect the internal state for all memory accesses and branch operations. First of all, we generate a set of arbitrary inputs for a chosen secret. These inputs can be either random (e.g. plain texts for encryption) or have a special structure with some random components (e.g. private keys or ephemeral secrets). We then execute the target binary on each input and log the following events:

	\begin{itemize}
		\item memory allocations
		\item branches, calls and returns
		\item memory reads and writes
		\item stack operations.
	\end{itemize}
	
	Absolute memory addresses may vary even for constant-time programs, e.g., due to ASLR and dynamic heap allocation. We use the trace of memory allocations and stack operations to compute relative memory addresses; our meta data then consists of a list of relative addresses for memory accesses and the branches to, from and within the code we are analyzing. Note that one can define other leakage sources based on the underlying microarchitecture and collect the state of relevant instructions for analysis, e.g., the \texttt{multiplication} on some ~\emph{ARM} platforms leaks information~\cite{armCortexM3}.
	
	\subsection{Preparing State Variables}
	We do not make any assumptions on the leakage granularity; compared to similar techniques, that stop at cache line level, we keep this parameter freely configurable. This has the advantage that the analysis can be restricted to leakage sizes that are actually relevant to the analyst: For example, as of writing this paper, on Intel processors the finest known attack has a leakage granularity of 4 bytes~\cite{moghimi2017memjam}. Applying our technique in 1-byte mode will give all positions where a leakage might occur, but if one only expects 4-byte leakages to be exploitable, this may yield some false positives. Instead, the security analyst can choose the leakage granularity that fits to the desired spatial resolution. After applying the chosen leakage granularity of $g\in\mathbb{N}$ bytes by discarding the lower $\log_2g$ bits of each address, we can acquire an efficient representation of a specific execution state by computing a hash value of all or a subset of the trace entries; a truly constant-time program should have identical hashes of the full trace for every secret input. If we are only interested in analyzing individual instructions, e.g., memory access leakages of a specific subroutine, we can as well just compute the hash for the subset of traces for a single instruction. 
	
	\subsection{Leakage Analysis}
	Our approach identifies any variations resulting from unique inputs and captured internal states per input. A naive approach is to compare the collected traces and divide them into classes. Observing more than one class informs us about secret-dependent operations. One can also compare raw traces sequentially which outlines all positions where the program behaves input-dependent and thereby allows to isolate the problematic sections. In addition to these simple approaches, we use MI to detect/locate these leakages, and to quantify the observable information.
	
	To simplify MI analysis, we assume that $X$ is a set of unique uniformly distributed input test cases, which trigger deterministic behavior of the investigated program. If the program makes use of randomization (e.g. blinding in RSA~\cite{kocher1996timing}), the test cases $x\in X$ should contain the corresponding sources of randomness too.
	
	Let $Y$ be a set of possible internal states (e.g. hashes of execution traces). We then define the execution state $T_i\subset X\times Y$ of the analyzed program at time point $i$ as
	\begin{equation*}
	(x,y)\in T_i\land(x,y')\in T_i\Rightarrow y=y',
	\end{equation*}
	i.e. each test case $x\in X$ appears at most once in $T_i$. The probability of one observed state $y\in Y$ is
	\begin{equation*}
	p_i(y)=\frac{\left|\{(x',y')\in T_i\,|\,y=y'\}\right|}{\left|T_i\right|}.
	\end{equation*}
	For the probability of pairs $(x,y)\in X\times Y$ we get
	\begin{equation*}
	p_i(x,y)=\left\{
	\begin{matrix}
	\frac{1}{\left|X\right|} & \text{if } (x,y)\in T_i,\\
	0 & \text{else,}
	\end{matrix}
	\right.
	\end{equation*}
	since each input and therefore each input/state tuple occur exactly once: $\left|T_i\right|=\left|X\right|$.
	
	With this knowledge we can finally compute the mutual information between test cases $X$ and the set of all occurring states $Y_i\isdef\{y\,|\,(x,y)\in T_i\}$:
	\begin{align*}
	I_i(X,Y_i)&=\sum_{(x,y)\in T_i}\frac{1}{\left|X\right|}\log_2\left(\frac{\frac{1}{\left|X\right|}}{\frac{1}{\left|X\right|}\cdot\frac{\left|\{(x',y')\in T_i\,|\,y=y'\}\right|}{\left|T_i\right|}}\right)\\
	&=\sum_{(x,y)\in T_i}\frac{1}{\left|X\right|}\log_2\left(\frac{\left|T_i\right|}{\left|\{(x',y')\in T_i\,|\,y=y'\}\right|}\right).
	\end{align*}

	\subsection{Interpretation of MI Score}
	As mentioned before, we can compute the MI for the entire trace, or a single instruction. A non-zero score for whole-trace MI tells us that an implementation has leakages, but it cannot locate the leakage point, and an implementation that has multiple leakage points over the execution period will have an aggregated MI value. The MI for single instructions is more precise, in which we can locate the instructions with positive score. The MI score $I_i(X_i,Y_i)$ is bounded by the amount of input bits $\log_2\left|X\right|$, and (for instruction MI) by the operand size: For example, an instruction that once accesses memory depending on 8 bits of the input will generate MI $\min\{8,\log_2\left|X\right|\}$. If we only execute $\left|X\right|=128$ test cases, we get MI score 7; for 256 or more test cases we get MI score 8.
	The analyzed MI score is an estimate of the average leakage over the given test cases. MI is the appropriate metric in cases where the analyzed inputs are not under the attackers control and commonly used in leakage quantification. Alternatively, the \emph{worst case} leakage for any attacker-chosen input is given by the \emph{min entropy}, which only considers the most likely guess. The use of min entropy instead of MI in \tool\ is recommended if the adversary has full control over the inputs and specific high-leakage inputs exist~\cite{minentropy2009}.

	\begin{figure*}[!t]
		\centering
		\subfloat{\includegraphics[width=.75\linewidth]{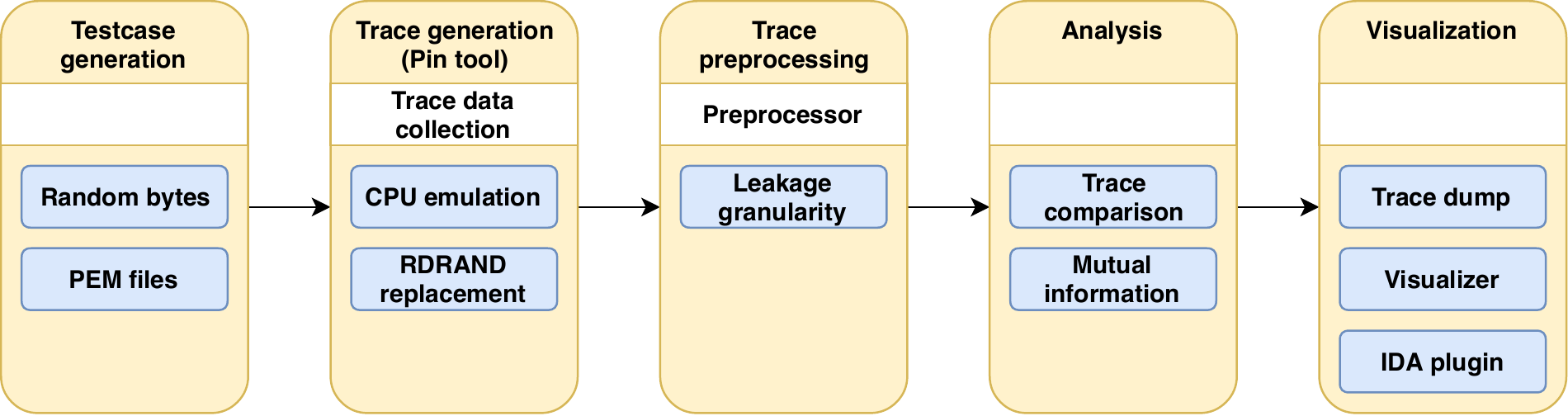}%
			\label{l}}
		\hfil
		\caption{The \tool~ pipeline: Given the software binary under test, the framework generates test cases using the selected source, that are then used to produce execution traces. These traces need to be preprocessed to extract important information. The resulting trace files can then be analyzed for leakages, which are shown to the user in the visualization stage. Each stage can be easily modified to add further functionality, that is used either interchangeably or in addition to existing features.}   
		\label{fig:pipeline}
	\end{figure*}

	\section{\tool~Framework}
	The \tool~framework is built as a pipeline with separate stages for test case generation, tracing and analysis (Figure \ref{fig:pipeline}). This modular design reduce the complexity, leading to easier extensibility: If one wants to implement additional analysis techniques apart from the ones that we already provide, she can directly add a new analysis stage, without needing to touch other parts like the trace generation. We will continue explaining these stages in more detail.
	
	\subsection{Investigated Binary}
	Although we are only interested in analyzing a specific function within a binary, we have to instrument and collect traces for the entire setup stage of the application before reaching to the analysis point, including the target library or executable itself, and parts of dependencies like system components. This process leads to an enormous decrease in analysis speed. A more efficient approach is to load and instrument the setup code only once, and then process the incoming test cases in a controlled loop. For libraries, we create a wrapper executable that executes an interface in a loop with new test cases. For executable applications, we can adopt in-memory fuzzing techniques where we inject hooks at the beginning and end of the target function and control the execution of the function to reset to the beginning with new test cases~\cite{bekrar2011finding}. To separate traces of the different test cases and avoid that the loop code causes false positives, we place calls to two instrumented dummy functions \texttt{PinNotifyTestcaseStart} and \texttt{PinNotifyTestcaseEnd}, which mark the start and the end of the analyzed section. A similar approach is taken by some fuzzers like \texttt{WinAFL}~\cite{WinAFL}, which use a built-in functionality of \texttt{DynamoRIO}~\cite{DynamoRIO} to exchange the argument list of \texttt{main} or a similar function.
	
	\subsection{Input Generation}
	The \tool~framework utilizes cryptographically secure pseudorandom number generators to create random test cases of any specified length. This performs well when analyzing cryptographic code, e.g. decrypting random ciphertexts. If a special input format is required, the test case generation code can be easily extended to produce such inputs, e.g. cryptographic keys in PEM format; this way, parts of the input can be kept constant while other parts are randomized, allowing to isolate the parameters which cause non-constant time behavior. Further, the framework supports passing a directory containing already generated inputs of any format.
	
	\subsection{Trace Generation}
	To trace the execution of individual test cases, we create a custom so-called \emph{Pintool}, which is a client library making use of \texttt{Intel Pin}'s dynamic instrumentation capabilities. In summary, our \emph{Pintool} logs the following events in a custom binary format on disk:
	\begin{itemize}
		\item module loads and the respective start and end addresses;
		\item calls to dummy functions in the instrumented executable, to identify start and end of a test case execution;
		\item sizes and addresses of allocated memory blocks through heap allocation functions such as \texttt{malloc} and \texttt{free} (Platform dependent), for resolving relative memory addresses;
		\item Stack pointer modifications, for resolving relative addresses;
		\item branches, calls and returns to and from all involved modules;
		\item memory reads and writes in investigated modules.
	\end{itemize}
	
	\subsubsection{Instruction Emulation} \label{sec:emul}
	Several cryptographic libraries use the \texttt{CPUID} instruction to detect the supported instructions for the respective processor and select a fitting implementation (that e.g. makes use of \texttt{AES-NI}). we enabled the \emph{Pintool} to change the output of this instruction. This allows to test arbitrary subsets of the instruction set that is available on the computer running \emph{Pin}.	
	
	As mentioned in~\autoref{sec:mi_technique}, cryptographic implementations might use randomization techniques like blinding to hide correlations between secret inputs and execution, or use ephemeral secrets. Some of these rely on the \texttt{RDRAND} instruction, which provides random numbers seeded with hardware entropy~\cite{RDRAND}. We provide an option to override the output of this instruction with arbitrary fixed values to control the randomization of the program under investigation.
	
	\subsection{Trace Preprocessing}
	The resulting raw trace files now need some preprocessing: First we add the common trace prefix that is generated before running the first test case, and which contains allocation data from the setup phase. In a second step, we calculate relative offsets of memory addresses. This involves associating branch targets with instruction offsets in the respective libraries, and identify offsets of memory accesses, such that traces generated using the same test case but during different runs of the \emph{Pintool} still match, regardless of the usage of randomized virtual addresses, e.g., \emph{ASLR}. For accesses to heap memory, we need to maintain a list of all currently allocated blocks: We use a stack to match the allocation size with their respective returned memory addresses, since in some implementations the heap allocator tends to call itself to reserve memory for internal bookkeeping. Finally the resulting preprocessed trace file is much smaller than the raw one (which can be discarded after this step), saving disk space and speeding up the following analysis stage.

	\subsubsection{Applying Leakage Granularity}
	We apply the leakage granularity immediately before the analysis starts; this way the preprocessed trace files are not modified, so the analysis can be performed on the same traces with different parameters. An analysis granularity of $g=2^b$ bytes ($b\in\mathbb{N}$) is introduced by discarding the $b$ least significant bits of each relative address.
	
	\subsection{Leakage Analysis}
	We implemented three different analysis methods in our framework: 
	
	\subsubsection{Analysis 1: Trace Comparison}
	The first analysis method implements the trace comparison technique; given two preprocessed traces, we compare them entry by entry to check whether they differ at all. This performs well for leakage detection of particularly small algorithms such as symmetric ciphers. Optionally, the user can use trace diffs to manually inspect varying sections.
	
	\subsubsection{Analysis 2: Whole-trace MI}
	For leakage detection of an entire logic and calculation of the average amount of input bits that might leak over arbitrary parts of the execution (assuming that the attacker has full access to the trace), we provide an option to estimate the MI between input data and resulting trace. Given a set $X$ of unique test cases, we need to determine matching outputs for each trace prefix. Since we can compute the final MI only after waiting for completion of all test cases, it would be inefficient to store the entire trace; instead we reduce the trace data by encoding information like relative memory accesses and branch targets into 64-bit integers, and then compress them into one 64-bit integer $y\in\{0,\ldots,2^{64}-1\}$ using a hash function. We store the resulting tuples of inputs and hashes in sets $T_i\subset X\times\{0,\ldots,2^{64}-1\}$ for each prefix length $i$. We then apply the methods from~\autoref{sec:mi_technique} to measure the trace leakage.
	
	\subsubsection{Analysis 3: Single-instruction MI}
	The average amount of bits leaked by a single memory instruction is calculated analogously to the trace prefixes: Here, for a specific instruction $i$, $T_i\subset X\times\{0,\ldots,2^{64}-1\}$ contains hashes of the accessed memory addresses for each input $x$. These hashes change when the accessed addresses, their amount or their order vary, thus we get the maximum amount of information that is leaked by the respective instruction.
	
	\subsection{Manual Inspection and Visualization}
	To be able to manually inspect the preprocessed traces, the program has an option to convert binary traces into a readable text representation. If \texttt{MAP} files with function names are available (exported by some compilers or disassemblers), these can be used to symbolize memory addresses. We also created an IDA python plugin to import our single-instruction MI results as disassembly annotations. This helps further analysis on which parts of functions and loops leak.
	
	Further we developed an experimental visualization tool, that renders function names and then draws an execution path. It also provides an option to render two traces simultaneously and highlight all sections where they have differences. This gives a quick overview of potential leakages and their structure.

	\section{Case Study I: Intel IPP}
	Intel's \emph{Integrated Performance Primitives (IPP)} cryptographic library aims to provide high performance cryptographic primitives that are compatible with various generations of Intel's processor~\cite{intelIPP}. \emph{Intel IPP} supports symmetric operations such as AES, as well as asymmetric signature and encryption schemes such as ECDSA. Intel IPP is used as the cryptographic backend for many of Intel's security products such as Intel SGX. Each of the implemented schemes in this library comes in variants optimized for different processors~\cite{inteldispatch}. The dynamic library checks the supported instruction set at runtime and chooses the most optimized implementation. However, developers can statically link toward a specific implementation by choosing the proper architecture code, e.g., \texttt{n8\_ippsAESInit} rather than \texttt{ippsAESInit}. In this case study, we test implementations for the variant optimized for processors supporting \emph{Intel® Advanced Vector Extensions 2 (Intel® AVX2)} with architecture code \texttt{l9}.
	
	\subsection{Applying \tool~MI Analysis to IPP}
	
	To be able to test \emph{Intel IPP} cryptographic implementations, we prepared wrappers that perform encryption and signing operations. For each tested implementation, we configured the wrappers for testing multiple test case scenarios: \textbf{1)} randomized plaintexts/ciphertexts to be encrypted/decrypted, or the message to be signed, \textbf{2)} randomized symmetric keys or private asymmetric keys, and \textbf{3)} random ephemeral secrets, when it is applicable, e.g., \emph{DSA} and \emph{ECDSA} as the input to MI Analysis. As suggested by chosen plaintext/ciphertext attacks, attacks on the cipher key and lattice attacks on ephemeral secrets~\cite{benger2014ooh}, using these scenarios, we are able to detect leakages that are dependent on various types of secrets. 
	
	\begin{table}[t!]
		\begin{center}
			\captionof{table}{Singe-instruction MI Analysis of Intel IPP cryptographic implementations v2018.2.185. All implementations are chosen from the \emph{l9} architecture code.} \label{tab:ippmi} 
			\begin{tabular}{ | p{1.2cm} | c | p{1.35cm} | p{1.05cm} | p{.9cm} | }
				\hline
				\textbf{\small{Scheme}} & \textbf{\small{Interfaces}} & \textbf{\small{Executed / Unique Instructions}} & \textbf{\small{Analysis Time (ms)}} & \textbf{\small{Leakage Found}}\\ \hline
				\small{3DES/ECB} & \makecell{\scriptsize{ippsDESInit}\\\scriptsize{ippsTDESDecryptECB}} & \small{$4074613$ / $70205$} & $11921$ & $0$ \\ \hline
				\small{SM4/ECB} & \makecell{\scriptsize{ippsSMS4Init}\\\scriptsize{ippsSMS4EncryptECB}} & \small{$4085517$ / $68221$} & $10004$ & $0$ \\ \hline
				\small{AES/CTR} & \makecell{\scriptsize{ippsAESInit}\\\scriptsize{ippsAESEncryptCTR}} & \small{$2138799$ / $49181$} & $27289$ & $2$ \\ \hline
				\small{DSA}~\scriptsize{(512)} & \makecell{\scriptsize{ippsDLPGenKeyPair}\\\scriptsize{ippsDLPSignDSA}} & \small{$12245281$ / $57423$} & $1735153$ & $2$ \\ \hline
				\small{RSA}~\scriptsize{(512)} & \makecell{\scriptsize{ippsRSA\_Decrypt}} & \small{$43987943$ / $55167$} & $275090$ & $1$ \\ \hline
				\small{ECDSA} \scriptsize{(SECP256R1)} & \makecell{\scriptsize{ippsECCPGenKeyPair}\\\scriptsize{ippsECCPSignDSA}} & \small{$4085155$ / $63785$} & $358373$ & $3$ \\ \hline
				\small{ECDSA} \scriptsize{(BN256)} & \makecell{\scriptsize{ippsECCPGenKeyPair}\\\scriptsize{ippsECCPSignDSA}} & \small{$5383210$ / $63699$} & $750188$ & \footnotesize{*} \\ \hline
				\small{ECDSA} \scriptsize{(SM2)} & \makecell{\scriptsize{ippsECCPGenKeyPair}\\\scriptsize{ippsECCPSignDSA}} & \small{$5158607$ / $63741$} & $353435$ & \footnotesize{*} \\ \hline
				\small{ECNR} \scriptsize{(SECP256R1)} & \makecell{\scriptsize{ippsECCPGenKeyPair}\\\scriptsize{ippsECCPSignNR}} & \small{$4028592$ / $62447$} & $281937$ & $2$ \\ \hline
				\small{SM2} & \makecell{\scriptsize{ippsECCPSignSM2}} & \small{$6021005$ / $64273$} & $554035$ & $3$ \\
				\hline\hline
				\multicolumn{2}{|c|}{\textbf{\small{Total}}}  & \small{$91208722$ / $618142$} & $73$~\scriptsize{minutes} & $13$ \\ \hline
				\multicolumn{4}{l}{\footnotesize{* Different curves did not change the results for ECDSA.}}
				
			\end{tabular}
		\end{center}
	\end{table}
	
	\autoref{tab:ippmi} shows the single-instruction MI analysis results, where symmetric ciphers: \emph{(Triple) DES}, \emph{AES} and \emph{SM4} and asymmetric ciphers: \emph{DSA}, \emph{RSA}, \emph{ECDSA}, \emph{ECNR} and \emph{SM2} have been tested. On our analysis setup, the total computational time to analyze $10$ different implementations with about $92$ million total instructions is $73$ minutes of CPU time, highlighting the efficiency of our method. Note that we performed analysis with input size $2^7=128$ (7-bit \emph{MI}) and input size $2^{10}=1024$ (10-bit \emph{MI}), for analysis of symmetric and asymmetric operations respectively. Although analysis with more iterations is possible, state-of-the art side-channel attacks on these implementations suggest that the random secret should show leakage behavior after this number of iterations. \emph{Intel IPP} uses two separate interfaces for the key schedule, and ephemeral secret generation for most implementations (\autoref{tab:ippmi}).

	\emph{(Triple) DES}, \emph{AES} and \emph{SM4} are block ciphers that use table-based S-Box operations. The results suggest that these implementations are heavily protected against memory-based leakages. Our target architecture code uses the \texttt{AES-NI} instruction set for \emph{AES} and \emph{SM4} operations. \texttt{AES-NI} is inherently secure against known attacks. However, testing the \emph{CTR mode} reveals some leakages. All asymmetric ciphers suffer from at least one leakage. For schemes that are based on elliptic curves such as \emph{ECDSA}, \emph{ECNR} and \emph{SM2}, Intel IPP supports various standard curves. As some developers optimize curve arithmetic differently for various standard elliptic curves, we tested the \emph{ECDSA} signing operation with three different curves: \texttt{SECP256R1}, \texttt{BN256} and \texttt{SM2}. However, the \emph{MI} analysis results are exactly the same for different choices of elliptic curves. We found a total of $13$ leakages in \emph{Intel IPP}, while some of these leakages are triggered through calling the same subroutine, e.g., both \emph{ECDSA} and \emph{SM2} use the leaky subroutine for \texttt{scalar multiplication}. We will discuss these subroutines in more detail.
	
	\subsection{Discovered leakages in Intel IPP}
	
	We have found $7$ different subroutines that have leakages, i.e., perform data-depended memory accesses or branch decisions (\autoref{tab:ippleakage}). We performed an initial analysis of these leakages using our visualization tool and \emph{IDA Pro}. The subroutine \texttt{gfec\_MulBasePoint} performs scalar multiplication of a scalar and point on the elliptic curve, as a common operation in all curve-based signature schemes: \emph{ECDSA}, \emph{ECNR}, \emph{SM2}. As defined by the signing algorithms~\autoref{sec:signing}, \emph{gfec\_MulBasePoint} leaks information about the ephemeral secret. This leakage occurs due to the dependability of the number of times the window-based multiplier loop processes the ephemeral secret. Further leakages exist in the curve operations after the scalar multiplication: The subroutine \emph{alm\_mont\_inv} leaks information during the mapping of $x$ coordinate of computed public point. As $(x_1, y_1)$ are not secrets in the signing operation, this leakage is not critical, and we refrain from further root cause analysis. Similarly, the subroutine \emph{cpModInv} has leakages with a relatively high MI score that is due to the secret-dependent loop count. \emph{cpModInv} performs a modular inversion operation using Extended Euclidean Algorithm (EEA). In \emph{ECDSA}, $k^{-1}$ leaks information about the secret ephemeral, and in \emph{SM2}, $(1+d_A)^{-1}$ leaks information about the secret signing key. \emph{ECNR} does not perform any modular inversion and is safe from leakages due to this subroutine. The existing leakage in \emph{cpModInv} subroutine also applies to \emph{DSA} where a modular inversion on ephemeral secret, $k^{-1}$ can leak.

	\begin{table}[t!]
		\begin{center}
			\captionof{table}{Discovered leakage subroutines within Intel IPP cryptographic implementations v2018.2.185. Some of the subroutines expose critical and potentially exploitable leakages.} \label{tab:ippleakage} 
			\begin{tabular}{ | l | p{1.4cm} | l | l | }
				\hline
				\textbf{\small{Subroutine}} & \textbf{\small{Affected}} & \textbf{\small{MI}} & \textbf{\small{Leakage Source}} \\ \hline
				\small{gfec\_MulBasePoint}& \small{ECDSA, ECNR, SM2}& \small{$0.86$ / $10$}  & \small{Conditional Loop} \\ \hline
				\small{cpMontExpBin}& \small{DSA}& \small{$3.73$ / $10$}  & \small{Conditional Loop} \\ \hline
				\small{cpModInv}& \small{DSA, SM2, ECDSA}& \small{$3.88$ / $10$}  & \small{Conditional Loop} \\ \hline
				\small{ExpandRijndaelKey}& \small{AES/CTR}& \small{$7.00$ / $7$}  & \small{Memory Lookup} \\ \hline
				\small{ippsAESEncryptCTR}& \small{AES/CTR}& \small{$0.13$ / $7$}  & \small{Conditional Loop} \\ \hline
				\small{gsMontExpWin}& \small{RSA}& \makecell{\small{$1.12$ / $10$}\\\small{$3.11$ / $10$}}   & \makecell{\small{Conditional Loop}\\\small{Memory Lookup}} \\\hline
				\small{alm\_mont\_inv}&\small{ECDSA, ECNR, SM2}& \makecell{\small{$5.33$ / $10$}\\\small{$9.98$ / $10$} } &  \makecell{\small{Conditional Loop}\\\small{Memory Lookup}} \\ \hline
			\end{tabular}
		\end{center}
	\end{table}
	
	\emph{Intel IPP} supports two distinct functions for performing Montgomery exponentiation. Exponentiation of big numbers is a common operation in schemes such as \emph{RSA} and \emph{DSA}. The \emph{RSA} algorithm uses the \emph{gsMonthExpWin} subroutine which is a window-based implementation of the Montgomery exponentiation. This function has leakages based on both memory lookup and conditional loop. The second Montgomery exponentiation subroutine~\texttt{cpMontExpBin} is a protected binary implementation that has leakage due to the conditional loop count. \emph{DSA} uses the latter, which leaks information about the ephemeral secret during computation of $(g^k \bmod p)$.

	The only leakage exposed during testing of symmetric ciphers are due to \emph{AES} key generation subroutine \texttt{ExpandRijndaelKey}, and calculation of the nonce length in \emph{CTR mode}. \texttt{ExpandRijndaelKey} is called every time the \texttt{ippsAESInit} is used. As the high \emph{MI} score shows, \emph{AES} key schedule used during the CTR mode has full leakage. This leakage can be considered critical in scenarios such as the SGX environment where an adversary has a high resolution side channel~\cite{moghimi2017cachezoom,wang2017leaky}. When the symmetric key is passed to the AES key schedule, a high resolution adversary can steal the secret key before any encryption/decryption. While \emph{AES/CTR} encryption uses \texttt{AES-NI}, there is a loop within this implementation where calculating the length of nonce leaks about the leading zero bits.
	
	\subsubsection{Leakage of Scalar Multiplication}
	Scalar multiplication in \emph{Intel IPP} uses a fixed-window algorithm with a window size of $5$: for a 256-bit ephemeral secret, as defined by \texttt{SECP256R1}, the algorithm performs $51$ iterations of the window operation. However, our dynamic analysis of the algorithm with various random ephemeral secrets shows that \texttt{gfec\_MulBasePoint} skips the leading zero bits and applies fewer windows if there are leading zero bits in the beginning, as the multiple of the window size. In this case, the main loop performs $50$ times for $2$, $49$ for $7$ and $48$ times for $12$, etc, leading zero bits. \emph{CacheQuote}~\cite{dall2018cachequote} exploits a similar vulnerability used by \emph{Intel EPID} signature scheme, but \emph{EPID} uses a different function of \emph{Intel IPP} for scalar multiplication~\texttt{cpEcGFpMulPoint}. As our discovery suggests, this was a common issue in \emph{Intel IPP} that was existed among other curve implementations. Although this implementation has countermeasure based on Scatter-Gather technique~\cite{brickell2006mitigating}, this vulnerability can easily be exploited in high resolution settings using a lattice attack~\cite{dall2018cachequote}.

	\begin{algorithm}[t]
		\caption{Bitmasked Montgomery Exponentiation}
		\label{alg:montexp}
		\begin{algorithmic}[1]
			\Procedure{BinExp}{base $g$, exponent $k$}
			\State $A \gets R \bmod p$
			\State $\widetilde{g} = \mathrm{MontMul}(g, R^2 \bmod p)$
			\State $m \gets 0$
			\State $i = 1$
			\While{$i < (\mathrm{BitLength}(k) \bmod 64)$} 
			\State $t \gets A \BitAnd \BitNeg m \BitOr \widetilde{g} \BitAnd {m}$
			\State $A \gets \mathrm{MontMul}(A, t)$
			\State $m = \BitNeg m \BitAnd k_i$
			\State $i = i + 1 - m$
			\EndWhile
			\For{$j \gets 1$ \textbf{to} $\mathrm{BitLength}(k) / 64$} 
			\State perform the same operations as above.
			\EndFor
			\State \Return $A$
			\EndProcedure
		\end{algorithmic}
	\end{algorithm}
	
	\subsubsection{Bitmasked Montgomery Exponentiation}
	The Montgomery exponentiation in \emph{Intel IPP} follows a bit-by-bit operation based on the Montgomery Reduction technique~\cite{ha1998common}. However, the implementation is protected by obfuscating the conditional statements as bit-masked operations. Therefore, the subroutine always executes the same \texttt{Montgomery multiplication (MontMul)} subroutine disregarding the value of the exponent bits. However, the exponent bits are used as a mask to choose the operand of the \texttt{MontMul} and to execute the \texttt{MontMul} two times with two different operands when the exponent bit is one.	Although this implementation looks secure at first sight, the exponent bits are used \textbf{1)} to calculate the exponent bit length, i.e., leading zero-bit leakage, and \textbf{2)} to decide the number of iterations of the loop. Based on Algorithm~\ref{alg:montexp}, the main loop executes two times if an exponent bit is one and once if the exponent bit is zero. This leaks the Hamming weight of the ephemeral secret to a microarchitectural adversary.

	Further, the algorithm performs a similar operation with separate instructions for different parts of the key. For example, for a 160-bit DSA exponent, the algorithm first processes the first 32 bits, and then another code section processes the remaining 128 bits of exponent. This gives an adversary a local Hamming weight leakage of the first 32-bit of the secret exponent. 
	
	\section{Case Study II: Microsoft CNG}
	The \emph{Cryptography API: Next Generation (CNG)} is the cryptography platform supplied with every Windows system beginning with Windows Vista, and replaces the older \emph{CryptoAPI} as the default cryptographic stack. It includes many common algorithms, including \emph{RSA}, \emph{AES}, \emph{ECDSA}. While the public API for \emph{Microsoft CNG} resides in the \texttt{BCrypt.dll} system file, its cryptographic implementations themselves are located in another library file, \texttt{BCryptPrimitives.dll}. Microsoft does provide neither source code nor documentation for the internal functionality, but one can download \texttt{PDB} symbol files from Microsoft's symbol server, which contain most of the internal function names, helping to reduce the reverse engineering effort.
	
	\subsection{Applying \tool~MI Analysis to CNG}
	As we did with IPP, we again created wrapper executables to call the respective library functions of \emph{RSA}, \emph{DSA}, \emph{ECDSA} and \emph{AES/ECB}. For \emph{AES}, the library uses the \texttt{CPUID} instruction to choose between two different implementations, one that uses \texttt{AES-NI} vector instructions, and a plain T-table based implementation. We tested both implementations by emulating the \texttt{CPUID} instruction, as explained in~\autoref{sec:emul}. The results are shown in~\autoref{tab:cngmi}. We analyzed a total of $21$ million instructions in $31$ minutes of CPU time, finding four different leakage points. For \emph{RSA}, we discovered that Microsoft's implementation behaves truly constant-time. \emph{ECDSA} and \emph{DSA} implementations both suffer from leakage due to calling the same subroutine for modular inversion.

	\begin{table}[t!]
		\begin{center}
			\captionof{table}{Singe-instruction MI Analysis of some of the bcryptprimitives.dll v10.0.17134.1 cryptographic implementations. }         \label{tab:cngmi} 
			\begin{tabular}{ | p{.95cm} | p{2.4cm} | p{1.4cm} | p{.95cm} | p{.85cm} | }
				\hline
				\textbf{\small{Scheme}} & \textbf{\small{Interfaces}} & \textbf{\small{Executed / Unique Instructions}} & \textbf{\small{Analysis Time (ms)}} & \textbf{\small{Leakage Found}}\\ \hline
				\small{AES/ECB} & \scriptsize{SymCryptAesEcbEncrypt} & \small{$2384298$ / $55451$} & $17546$ & $0$ \\ \hline
				\small{AES/ECB} & \scriptsize{SymCryptAesEcbEncryptAsm} & \small{$2324391$ / $63179$} & $26211$ & $2$ \\ \hline
				\small{DSA}~\scriptsize{(512)} & \scriptsize{MSCryptDsaSignHash} & \small{$3586162$ / $63748$} & $223356$ & $1$ \\ \hline
				\small{RSA}~\scriptsize{(1024)} & \scriptsize{MSCryptRsaDecrypt} & \small{$8073605$ / $66454$} & $760450$ & $0$ \\ \hline
				\small{ECDSA} \scriptsize{(SECP256R1)} & \scriptsize{MSCryptEcDsaSignHash} & \small{$4764783$ / $64732$} & $831136$ & $1$ \\ \hline\hline
				\multicolumn{2}{|c|}{\textbf{\small{Total}}}  & \small{$21133239$ / $313564$} & $31$~\scriptsize{minutes} & $4$ \\
				\hline
			\end{tabular}
		\end{center}
	\end{table}
	
	\subsection{Discovered leakages in Microsoft CNG}
	Analyzing the aforementioned algorithms yielded two leakage candidates (see Table \ref{tab:cngleakage}); the first one resides within the modular inversion function of \emph{DSA} and \emph{ECDSA} and is used for all processors. The MI returns full leakage for the modular inversion leakage, implying that the implementation is heavily unprotected. The second one is in the encryption function of \emph{AES} and only used by processors not supporting \texttt{AES-NI}. As it is a table-based implementation, the leakage is expected.

	\begin{table}[t!]
		\begin{center}
			\captionof{table}{Discovered leakage subroutines within bcryptprimitives.dll v10.0.17134.1 cryptographic implementations.} \label{tab:cngleakage} 
			\begin{tabular}{ | l | p{1.0cm} | p{0.5cm} | p{1.3cm} | }
				\hline
				\textbf{\small{Subroutine}} & \textbf{\small{Affected}} & \textbf{\small{MI}} & \textbf{\small{Leakage Source}} \\ \hline
				\small{SymCryptFdefModInvGeneric}& \small{DSA, ECDSA}& \small{$10.00$ / $10$}  & \small{Conditional Loop} \\ \hline

				\small{SymCryptAesEncryptAsmInternal}& \small{AES}& \small{$7.96$ / $10$}  & \small{Memory Lookup} \\ \hline
			\end{tabular}
		\end{center}
	\end{table}

	\subsubsection{Leakage of Modular Inversion}
	The modular inversion function that is used for \emph{DSA} and \emph{ECDSA} gives full \emph{MI} on 1024 signing operations for random ephemeral secrets with fixed key and plaintext. This subroutine does not have any constant-time protection. However, while this is a non-constant time behavior and suggests that the ephemeral leaks, we considered this as not exploitable; Microsoft protects this implementation through a masking countermeasure. The masking countermeasure for modular inversion works as follow:
	\begin{enumerate}
		\item A mask value $m$ is generated randomly.
		\item The ephemeral secret $k$ is multiplied by $m$ before the modular inversion: $s = (k\cdot m)^{-1}(z + x) \bmod q$
		\item Then the signature $s$ is multiplied again with $m$ to produce the correct signature: 
		\begin{gather*}
		s = sm = (k\cdot m)^{-1}(z + x)\cdot m 
		= k^{-1}(z + x)
		\end{gather*}
	\end{enumerate}
	Thus, the implementation leaks $k\cdot m$, where $m$ is a random per-signature generated mask, effectively preventing extraction of useful information. Leakage of ephemeral keys is exploitable~\cite{benger2014ooh}, the randomized product of ephemeral key and a random value is not.
	
	\subsubsection{Leakage of AES T-table Lookup}
	The non-vector version of \emph{AES} uses a common lookup table implementation, where four so-called \emph{T-tables} combine the steps \texttt{SubBytes}, \texttt{ShiftRows} and \texttt{MixColumns}. Each round consists of four of such lookups per table, leading to $16\cdot r$ memory accesses per encryption, where $r\in\{10,12,14\}$ is the number of rounds. The 8-bit indices used for the table accesses depend on the plaintext and the key; since the \emph{MI} is $7.96$ for $1024$ measurements, these indices can be considered fully leaking. Each table entry has $4$ bytes size, thus each T-table has $1024$ bytes, and therefore takes $16$ cache lines on an Intel processor; such implementations have already been shown to be exploitable with cache attacks \cite{bernstein2005cache}.

	\section{Related Work}
	\textbf{Programming languages} can support constant-time code generation and verification~\cite{sinha2017compiler,cauligi2017fact,bond2017vale}. The general approach is to support annotation of security-critical variables and to generate instructions that operate obliviously on annotated secrets. Annotated secrets can be verified for constant-time behavior using SMT-based techniques~\cite{bond2017vale}. Constant-time behavior can be enforced for some operations by using primitives such as oblivious RAM (ORAM)~\cite{stefanov2013path} and obfuscated execution~\cite{rane2015raccoon}. Language-based approaches are not widely used, and annotation is an error-prone task. 
	
	\textbf{Black-box testing} approaches use statistical methods to quantify leakages of physical channels~\cite{coron2000statistics}. In particular, \emph{Dudect}~\cite{reparaz2017dude} performs black-box timing analysis, in which the timing of a target system with different inputs will be analyzed using the \emph{t-test}~\cite{welch1947generalization}, but these black-box techniques do not scale to microarchitectural attacks with a gray-box model. With an abstract model of the leakage channel, methods based on~\textbf{Static Program Analysis} are proposed to analyze program code and to quantify leakages~\cite{kopf2012automatic,almeida2013formal,barthe2014system,almeida2016verifying,blazy2017verifying}. Similar to language-based approaches, these techniques are limited to correct annotation of the source code. While some of these approaches are limited to the source code and cannot find leakages that are potentially introduced by the compiler~\cite{almeida2013formal}, others perform the analysis on the lower level LLVM bitcode~\cite{almeida2016verifying} or the annotated machine code~\cite{barthe2014system,blazy2017verifying}. However, they rely on the availability of the source code. \emph{CacheAudit}~\cite{doychev2015cacheaudit,doychev2017rigorous} is based on Static Binary Analysis (SBA). SBA approaches need to initially reconstruct the original basic blocks and control flow graph. Precise reconstruction of the program semantic and control flow graph is infeasible without the runtime information, by just using static disassembly~\cite{andriesse2016depth}. As a result, while they give formal guarantees on the absence of leakages, they do not scale to accurately analyze large program binaries, e.g., \emph{CacheAudit} approach has only been tested on rather simple algorithms such as sorting and symmetric encryption. Other proposals based on \textbf{Symbolic execution} quantify side-channel leakage by determining symbolic secret inputs that affect the runtime behavior~\cite{pasareanu2016multi,chattopadhyay2017quantifying}. However symbolic execution is an expensive approach, and the proposed methods require access to the source code.
	
	In this work, we leverage \textbf{Dynamic Program Analysis} techniques to accurately locate microarchitectural leakage in software binaries, as they execute on the processor. \emph{ctgrind}~\cite{langley2010ctgrind} based on LLVM memcheck can check all branches and memory accesses to make sure that they do not have dependency on secret data. Irazoqui et al.\ instrument the source code to obtain and analyze cache traces using MI~\cite{Irazoqui16Leak}. Sensitive code sections are identified by taint analysis. On binary-only approaches, \emph{CacheD}~\cite{wang2017cached} analyzes binaries based on symbolic execution and constraint solving. They initially use DBI to get execution traces for a set of input values; then, given the information which input values are considered secret, a taint analysis extracts all instructions that work with secrets, either directly or indirectly. These instructions are then analyzed using symbolic execution to detect whether cache leakages exist. In comparison, our method aims at maximum performance without too much loss of accuracy by only storing necessary information and using hash compression to get small execution states. The symbolic execution approach introduces a large bottleneck, as their analysis time suggest. This saving of computation time allows us to detect also other types of leakages like differing loop counts or byte-level memory access differences. Also, since \tool~is designed as a modular open source framework, one can implement arbitrary analysis stages for other types of leakages. Zankl et al.~\cite{zankl2016automated} use DBI to collect traces for instruction based leakage detection. They use \emph{t-tests} for leakage analysis and only test for execution flow leakages. \emph{STACCO}~\cite{xiao2017stacco} is focused on differential trace analysis for Bleichenbacher attacks ~\cite{bleichenbacher1998chosen}. Independently, \emph{DATA}~\cite{WeiserDATA} follows a similar approach based on DBI. They use trace differentiation and \emph{t-tests} for leakage analysis. As of our knowledge, our work is the first that has been tested on actual closed-source binaries.
	
	\section{Conclusion}
	The lack of efficient and practical tools for leakage analysis of binaries leave the reliability of these untested deployed implementations a mystery. To be able to analyze the compiler outputs and closed-source libraries, we have created an extensible framework that supports various types of microarchitectural leakages based on \texttt{instruction and data cache}, \texttt{MOB}, \texttt{BPU}, etc. \tool\ can be extended to analyze other and future side channels. Our framework leverages \emph{DBI} to collect the internal state of a program under test, and it applies multiple analysis techniques based on trace comparison and \emph{MI}. \tool~is open source and is publicly accessible: \url{https://github.com/UzL-ITS/Microwalk}. We used this framework to thoroughly analyze two widely used closed-source libraries, \emph{Intel IPP} and \emph{Microsoft CNG}. The tested implementations are optimized for the current generation of Intel processors. Our report shows that side-channel countermeasures for these implementations are still not fully leakage-free, e.g., all the curve-based signature schemes in \emph{Intel IPP} suffer from at least one vulnerability.  We have identified several leakages in symmetric and asymmetric ciphers, and reported them to the respective vendors. Our analysis shows that despite the existing efforts on protecting these implementations, some of them still suffer from security-critical leakages.
	
	\subsection{Future Work}
	\subsubsection{Coverage-based Fuzzing}
	We use random test cases to get a uniform random distribution of potential memory accesses and execution paths; while this works well with cryptographic implementation, it would not scale to targets such as protocols or data structures. Coverage-based \emph{Fuzzing}\cite{mcnally2012fuzzing} is a technique to generate test cases with the aim of achieving maximum code coverage; while it was originally developed to find software bugs, e.g., memory corruption, the same approach can be applied for finding side-channel leakages, e.g., leakage in the \emph{JPEG} library~\cite{xu2015controlled}. We have already implemented an experimental support for using \texttt{WinAFL}\cite{WinAFL} as a test case generator; in that setting \texttt{AFL} helps to generate samples with higher coverage, while at the same time the test cases are sent to our framework for further processing. It is desirable to enhance this experimental feature and apply it to non-cryptographic implementations that are critical in terms of side-channel security.

	\subsubsection{Distinguishing leakages in call graph}
	We observed that in some cases control flow leakages in the higher level algorithm residing at the top of the call chain hide leakages in the subroutines invoked in deeper levels. Also, if separate functions use a common subroutine, a positive MI result in this subroutine can not easily be assigned to its root cause. We therefore propose to add an option to \tool\ to take the call graph into account when computing mutual information.

	\medskip\noindent{\bf Responsible Disclosure} 
	We have informed the \emph{Intel Product Security Incident Response Team (PSIRT)} and \emph{Microsoft Security Response Center (MSRC)} of our findings. \emph{MSRC} has not responded. After the initial report, we noticed that Intel have already patched \texttt{gfec\_MulBasePoint} in Intel IPP v2018.3.240. Intel have acknowledged the receipt for the remaining vulnerabilities. Here is the time line for the responsible disclosure:
	
	\begin{itemize}
		\item \textbf{06/22/2018:} We informed our findings to the Intel Product Security Incident Response Team (Intel PSIRT) and the Microsoft Security Response Center.
		\item \textbf{06/25/2018:} Intel PSIRT acknowledged the receipt. 
		\item \textbf{07/31/2018:} Intel PSIRT confirmed a work-in-progress patch for IPP 2018 update 4 (CVE-2018-12155, CVE-2018-12156).
	\end{itemize}
	
	\medskip\noindent{\bf Acknowledgements} 
	This work is supported by the National Science Foundation, under grant CNS-1618837.

	\bibliographystyle{ACM-Reference-Format}

	\citestyle{acmnumeric}

	\bibliography{ref}

\end{document}